\documentstyle[prl,aps,multicol,epsf]{revtex}

\begin{document}
\draft

\title{Thermally-Activated Magnetic Reversal Induced by a
Spin-Polarized Current}

\author{E. B. Myers, F. J. Albert, J. C. Sankey, E. Bonet, R. A. Buhrman, and
D. C. Ralph}
\address{Cornell
University, Ithaca, NY 14853}
\date{\today}
\maketitle

\begin{abstract}

We have measured the statistical properties of magnetic reversal in
nanomagnets driven by a spin-polarized current.  Like reversal
induced by a magnetic field,
spin-transfer-driven reversal near room temperature exhibits the properties
of thermally-activated escape over an
effective barrier.  However, the spin-transfer effect produces
qualitatively different behaviors than an applied magnetic field.  We
discuss an effective current vs.\ field phase diagram.
If the current and field are tuned so that their effects oppose
one another, the magnet can be driven superparamagnetic.
\end{abstract}

\par
\pacs{PACS numbers: 73.40.-c, 75.60.Jk, 75.70.Pa}

\begin{multicols} {2}
\narrowtext

	A spin-polarized current traversing a magnetic multilayer
can, through exchange interactions, alter the orientation of
ferromagnetic moments, producing domain reversal or exciting
spin waves.  Following early predictions
\cite{berger,slon,bazaliy}, these spin-transfer effects have been observed
\cite{tsoi,sunexp,myers,wegrowe,katine,albert,tsoi2,grollier}
and they are
generating
interest as an alternative to the use of magnetic fields for
switching elements in magnetic memories. Competing theories of the
effect have thus far considered only the limit when the temperature
$T\!=\!0$.  The initial theories,
which we shall call the torque model
\cite{berger,slon,clear}, predict
that domain reversal should occur when spin transfer from the current
produces a torque that exceeds the magnetic damping.  The switching
current is therefore not determined by an energy barrier.
Alternative approaches have predicted that the spin-polarized
current may provide an effective magnetic field favoring parallel or
antiparallel alignment of adjacent magnetic layers, thereby altering
an energy barrier for reversal \cite{heide,zhang}.
Here we report, for measurements of spin-transfer-driven reversal near room
temperature, broad distributions of
switching currents that depend strongly on temperature, similar to
the familiar distributions of switching fields measured when an
applied magnetic field drives thermally-activated magnetic reversal
\cite{wernsdorfer,koch}.  This
indicates that the primary effect of spin transfer near room
temperature is to alter a
thermally-activated over-barrier switching process.
Nevertheless, by comparing the effects of applied currents and
applied magnetic fields, we demonstrate that the
spin-transfer effect cannot be understood as an effective magnetic
field favoring parallel or antiparallel magnetic alignment
\cite{heide,guittienne}.
If we extrapolate our switching
currents and fields to $T\!=\!0$, we find qualitative agreement with
the torque model.

	Figure 1(a) is a schematic of our device geometry.  The
fabrication process\cite{albert} employs electron beam lithography
and ion milling to form a
pillar with cross section ranging from $\sim 50 \times 50$ nm (sample 
1) to $\sim
130 \times 60$ nm (sample 3)
from a multilayer
of 80 nm Cu/40 nm Co/6 nm Cu/3 nm Co/10 nm Au. The milling step is timed so
\begin{figure}
\begin{center}
\leavevmode
\epsfxsize=7.7 cm
\epsfbox{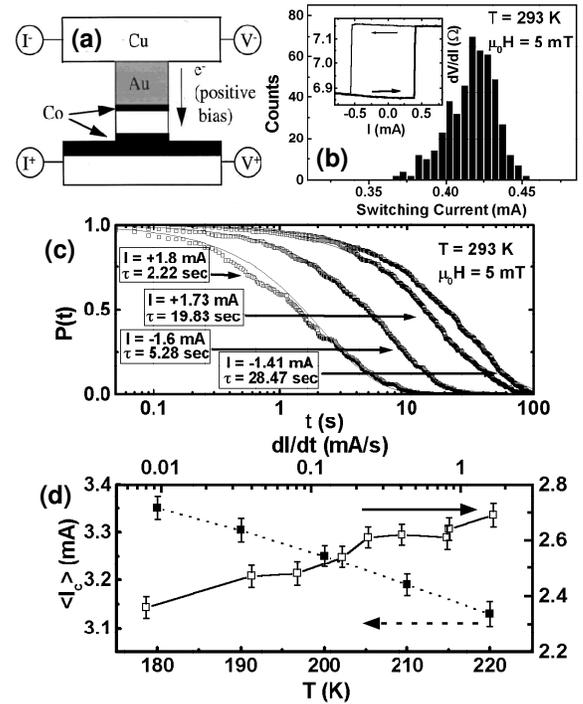}
\end{center}
\vspace{-0.1 in}
\caption{
\label{figure1}
(a) Schematic of the nanopillar device. (b inset) Differential resistance
vs.\ $I$ for device 1.
$H$ is
along the easy axis of the nanomagnet.
(b) Distribution of switching currents for device 1,
for the parallel to antiparallel transition.
(c) Waiting time distributions for device 2.
The distributions are fit to the function $e^{-t/\tau}$.
(d) Dependence of the mean switching current $\langle I_c \rangle$ on
$T$ (closed squares) and current-sweep rate at room
$T$ (open squares) for device 3 at $H\!=\!0$. The bars show
the width $2\sigma_{I_c}$ of the distributions.}
\end{figure}
\noindent
that the thicker
Co layer is left as
an extended film.
The differential resistance $dV/dI$
as a function of bias current $I$ perpendicular to the layers is
plotted in the inset to Fig.~1(b). If the moments of the two magnetic
layers are initially parallel (P) and the current is swept from
negative to positive\cite{signconv}, $dV/dI$ jumps at a critical
current $I_c^+$, where the moment
of the small Co nanomagnet is driven antiparallel (AP) to
the
thicker Co film. The device remains in this AP state
until the current is swept down past a negative value $I_c^-$, at
which point the spin-transfer effect drives the
nanomagnet back parallel to the thicker film. This asymmetry in $I$ is in
agreement with the theories
\cite{berger,slon,heide,zhang,waintal,xia,stiles}, and with previous
switching studies
\cite{myers,katine,albert,grollier}.

	The value of current at which the magnet reverses
varies from sweep to sweep.  A histogram of $I_c^+$ at room
$T$ is shown in Fig.~1(b) for a
current-sweep rate of $80~\mu$A/s.  Similar histograms are found for all of
the eight samples we have studied in detail.
The stochastic
nature of the switching is confirmed in
experiments in which we hold the sample at a
fixed $I$ near a switching threshold; there is a waiting time before
switching that also displays a broad distribution \cite{wernsdorfer}.
Probabilities $P(t)$ that the magnet has not
reversed in a time $t$ are plotted in Fig.~1(c) for a second sample
and compared to
exponential decay.
As a function of either decreasing $T$ or increasing
sweep rate $dI/dt$, the distributions of $I_c$ shift
strongly to larger values of $|I|$
(Fig.\ 1(d)).
At temperatures below 100-150 K,
the switching events generally consist of multiple jumps in
resistance rather than a single jump.  Related multiple jumps have been
observed previously
\cite{katine,albert}, and they indicate that the reversal
mechanism is not coherent rotation but rather a more
complicated process like domain wall nucleation
\cite{wernsdorfer96,koch2}.

Our main focus in this Letter will be on the statistical
distributions of switching currents and switching fields as both $I$
and magnetic field $H$ are applied to the sample.
By studying the interplay between spin transfer and $H$, we can gain new
insights into the mechanism underlying spin transfer. Fig.~2(a) shows
the mean values of the
current-induced reversal distribution as a function of constant $H$ at room $T$
measured with a current-sweep rate of $80 \mu$A/s.
$H$ for all data in the paper is applied within 5$^{\circ}$ of the
easy axis of the nanomagnet.
Both critical currents, $\langle I_c^+\rangle$ and
$\langle I_c^-\rangle$, shift towards
more positive $I$ as a function of $H$, with
the AP$\rightarrow$P transition $\langle I_c^-\rangle$ eventually
shifting more rapidly until it intersects
$\langle I_c^+\rangle$ and the nanomagnet is no longer
bistable \cite{albert}.  In Fig.~2(b) we show
the corresponding standard deviations for the switching currents,
$\sigma_{I_c}^+$ and $\sigma_{I_c}^-$.  These show unexpected differences.
At low fields, $\sigma_{I_c}^+$ and $\sigma_{I_c}^-$ are roughly equal to
one another. However, as $H$ increases near the values required for a
field-driven transition to the parallel state, $\sigma_{I_c}^-$ increases
by more than
threefold, while $\sigma_{I_c}^+$ decreases slightly.

If the roles of $I$ and $H$ are reversed, the results give a first
indication that spin transfer acts quite differently than
the magnetic field.
The inset to Fig.~2(c) shows a plot of $dV/dI$
near zero current bias versus $H$ along the easy axis for the same
device as that used for
Figs.~2(a) and 2(b).
The low-field transition to the higher-resistance state is due to the
reversal of the lower-coercivity extended Co film, while the
higher-field transition to low resistance corresponds to the
switching field
\begin{figure}
\begin{center}
\leavevmode
\epsfxsize=7.7 cm
\epsfbox{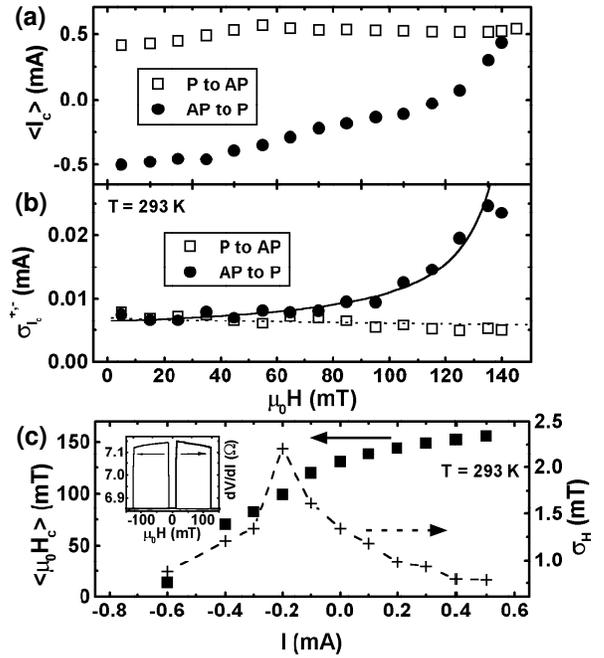}
\end{center}
\caption{
\label{figure2}
(a) Mean and (b) standard deviation of $I_c$ for
device 1
as a function of $H$.
Lines in (b) are fits of the standard deviations to Eq.~(4).
(c) Mean and standard deviation of $H_c$
as a function of $I$ for device 1. Inset: the low-bias
magnetoresistance loop.}
\end{figure}
\noindent
of the nanomagnet, $H_c$.
$\langle H_c \rangle$ and $\sigma_{H_c}$ as a function of
$I$ are shown in Fig.~2(c).
While $\langle  H_c \rangle$ monotonically increases
with $I$, there is a distinct peak in $\sigma_{H_c}$, occurring at
$-0.2$ mA.

The striking behavior exhibited in Fig.~2 can be
explained by extending the theory of thermally-activated
magnetic-field-induced switching.  We will employ the Kurkij\"{a}rvi
approach, in which the effective activation barrier to reversal is
tuned by varying an external parameter at a constant rate\cite{Kurkijarvi}.
For generality we will call the
barrier-reducing parameter $D$, and we will apply
the theory when $D\!=\!I$ as well as $D\!=\!H$.  The Kurkij\"{a}rvi
calculation assumes that the activation barrier has the approximate
form of a power law $U(D) = U_0(1-D/D_{c0})^{a_D}$ where $D_{c0}$ is
the parameter at which the barrier vanishes at $T=0$, and $a_D$ is a
constant.
The mean and standard deviation of the distributions for the
switching point $D_c$
are, to leading order\cite{garg},
\begin{equation}
\langle D_c\rangle = D_{c0} \left[ 1-\left(\frac{k_BT}{U_0}
A(T,\dot{D}) \right)^{1/a_D} \right],
\end{equation}
\begin{equation}
\sigma_{D_c} = \frac{|D_{c0}|}{a_D}
\left(\frac{k_BT}{U_0}\right)^{1/a_D}\left[A(T,\dot{D})\right]^{(1-a_D)/a_D},
\end{equation}
\begin{equation}
A(T,\dot{D}) = ln \left[\frac{1}{\tau_0a_D}
\frac{k_BT}{U_0} \frac{|D_{c0}|}{|\dot{D}|}
\left(\frac{ |D_{c0}|}{|D_{c0}-\langle D_c\rangle|}
\right)^{a_D-1}\right],
\end{equation}
where $\dot{D}$ is the sweep rate and $\tau_0$ is
the attempt time.

We should note
that, in addition to reducing the barrier to reversal, the 
application of current could also heat the device.  However, by
comparing the magnitude of $I$-dependent changes in DC resistance at
the relevant current levels (a few mA) to the $T$-dependence of the low-bias
resistance, we estimate
that the devices are heated by at most 2-3 K above room temperature.
This corresponds to a 1\% effect on
$\langle I_c \rangle$, which can be
neglected in our analysis.

If we apply a fixed magnetic field while sweeping $I$, $H$ has two
effects on the $I_c$
distributions. First, $H$
alters the effective zero-current barrier height $U_0$.  For the
AP$\rightarrow$P transition, the barrier is reduced from $U_0$ to
$U_0(1-H/H_{c0})^{a_H}$, while
the P$\rightarrow$AP barrier is increased to $U_0(1+H/H_{c0})^{a_H}$.
Second, the magnetic field can modify the zero-temperature critical
currents
$I_{c0}^{\pm}$ for the two transitions; the form will depend on the
microscopic model.
However, in the models proposed to date\cite{berger,slon,heide}, this
dependence is linear, so that we will take $I_{c0}^{\pm}(H) =
I_{c0}^{\pm}(0) (1 - H/H_s^{\pm})$, where $H_s^{\pm}$ are model-dependent.
Inserting these two quantities into Eq.~(2) and neglecting the weak
$H$-dependence of A yields
\begin{equation}
\sigma_{I_c}^{\pm}(H) \propto (1 - H/H_s^{\pm})/(1 \pm H/H_{c0})^{a_H/a_I}.
\end{equation}
If spin transfer merely acted as an additional effective field in the direction
of $H$,
then we should have $a_I = a_H$ and
$H_s^{\pm} = \mp H_{c0}$.   In this case the numerator and
denominator in Eq.~(4) cancel and $\sigma_{I_c}^{\pm}$ should be
$H$-independent.  This does not describe the data. In contrast, within
Slonczewski's torque model, $H_{c0}$ and $|H_s^{\pm}|$ differ.
For a thin-film nanomagnet, $H_{c0}$ is
set
by the small in-plane anisotropy $H_{coercive} \approx 150$ mT, while
the field intercept $|H_s^{\pm}| \approx H_{coercive}+H_{demag}$,
where $H_{demag}$ represents the additional
effect of the demagnetizing field
as the moment precesses out-of-plane (for single-domain magnets
undergoing coherent rotation, $\mu_0 H_{demag} =  \mu_0 M/2 \approx 
850$ mT for Co)
\cite{katine,suntheory}.
As a result, within this approach $\sigma_{I_c}^-$ diverges
at $H = H_{c0}$, while
$\sigma_{I_c}^+$ slowly decreases, in excellent agreement with the
data.  The lines in Fig.~2(c) illustrate the results of this
model using
$\mu_0 H_{c0}$ = 150 mT, $|\mu_0 H_s^{\pm}|$ = 230 mT, $a_I = a_H$,
and the
scale factors $\sigma_{I_c}^+(H=0) = 0.007$ mA, $\sigma_{I_c}^-(H=0) =
0.0065$ mA.  The fact that $|H_s^{\pm}|$ is less than $M/2$
may be due to non-single-domain dynamics.

The dependence of $\sigma_{H_c}$ on $I$
can be understood within the same model by considering
the nature of the $T\!=\!0$ stability boundaries for
P and AP alignment. A simple
relation allows us to connect the measured histogram means and widths to the
$T\!=\!0$ stability boundaries: from
Eqs.~(1)-(3),
since the function $A$ depends weakly on its 
\vspace{-.2 in}
\begin{figure}
\begin{center}
\leavevmode
\epsfxsize=7.7 cm
\epsfbox{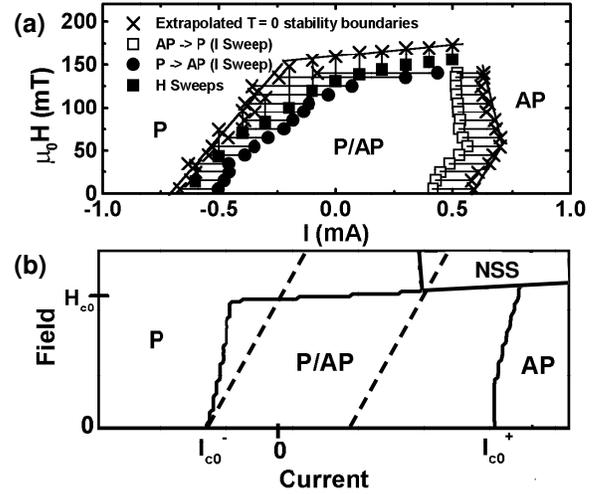}
\end{center}
\vspace{-0.1 in}
\caption{
\label{figure3}
(a) $T\!=\!0$ current-field stability diagram (crosses) for device 1,
showing the parallel (P), antiparallel (AP), and
bistable (P/AP) regimes, determined as described in the text. The
circles and squares are the same data as in Fig.~2, and solid lines
are guides to the eye.
(b) Solid lines: Form of the stability diagram in the Slonczewski
torque model for a single domain magnet with easy-plane
and in-plane anisotropies.  NSS is a regime where non-static states exist.
The critical currents $I_{c0}^{\pm}$ are related
through $I_{c0}^+/I_{c0}^- = g(\pi)/g(0)$, where g is the Slonczewski
polarization factor [2].
Dotted lines: Stability boundaries in the effective-field model.}
\end{figure}
\noindent
variables, to a good approximation
$\sigma_{I_c}^{\pm} \propto |I_{c0} - \langle I_c^{\pm} \rangle|$ and
$\sigma_{H_c} \propto H_{c0} - \langle H_c \rangle$. We can estimate the
proportionality constant self-consistently as follows. The normalized 
sweep rate
$\frac{|\dot{D}|}{|D_{c0}|}
\sim$ 0.05--0.1~s$^{-1}$ and we use an attempt time
$\tau_0\approx 100$ ns \cite{koch,tcomment}.   By fitting
the dependence of $\langle I_c \rangle$ on sweep rate for this sample
to Eq.\ (1)-(3), assuming
the barrier exponents to be $a_I = a_H \approx$ 1.5\cite{victora},
we find an $H\!=\!0$ effective barrier of
$U_0 =$ 1.5 - 2 eV. Inserting these values
into Eq.~(1)-(3) yields a proportionality constant $\sim 0.1$, with
the dominant uncertainty associated with $\tau_0$.
The resulting estimates for the $T\!=\!0$ stability boundaries,
extrapolated from the room-$T$ measurements of
$\langle I_c^{\pm} \rangle$ and $\langle H_c \rangle$,
are marked by the
crosses in Fig.~3(a).
The maximum in $\sigma_{H_c}$ is associated with
the knee in the stability diagram where the critical-current line
$I_{c0}^-(H)$ joins the critical-field line $H_{c0}(I)$.  In this
region the magnet
is maximally subject to thermally-activated reversal, through the 
combined effects of $I$ and $H$, and therefore
$\sigma_{H_c}(I)$ is a maximum. If spin transfer
worked as an effective field,
$I_{c0}^-(H)$ and $H_{c0}(I)$ would fall on one line, and there would
be no maximum in $\sigma_{H_c}(I)$.

In order to compare these results to the torque model, we have calculated
the stability boundaries
within the $T\!=\!0$ Slonczewski picture by numerically integrating
the Landau-Lifschitz-Gilbert equation with a spin-torque term
for a single-domain magnet with easy-plane and in-plane anisotropies
(Fig.~3(b)).
Although we do not expect this model to be quantitatively
accurate if the rever-
\begin{figure}
\begin{center}
\leavevmode
\epsfxsize=7.7 cm
\epsfbox{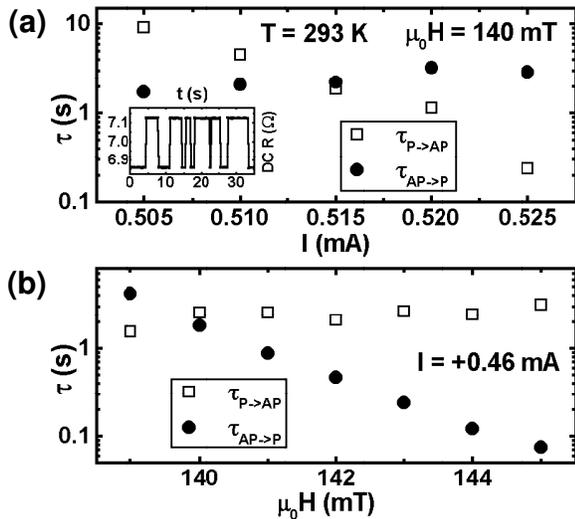}
\end{center}
\vspace{-0.1 in}
\caption{
\label{figure4}
Inset: DC resistance versus time for device 1 with
$\mu_0 H=140$ mT and $I=+0.51$ mA at room $T$.  (a,b) Mean
switching times for
the two resistance states. The time scales in (a) and (b)
do not match precisely because of a
small shift in the sample between measurements.}
\end{figure}
\noindent
sal mechanism is not single-domain,
it has the qualitative 
features needed to understand the data.
It predicts that the $I_{c0}^-(H)$ and $H_{c0}(I)$ lines are distinct, and
intersect at
a knee located at a negative value of $I$.

If we bias the sample near the point $\mu_0 H=140$ mT and $I=+0.5$
mA where the spin-transfer effect and $H$ oppose each other,
we observe telegraph-noise-type switching between resistance states
(inset, Fig.~4(a))\cite{tsoi}.
Unlike previous telegraph-noise studies in nanomagnets\cite{wernsdorfer},
which were done by
applying $H$ perpendicular to the easy axis so that
the moment jumped
between two closely-separated angles,
the jump here is between approximately full
P and AP alignment.  Most remarkably, the mean switching times for
the two types of
transitions depend very differently on $H$ and $I$.
If $H$ is held fixed along the easy axis and $I$ is increased
(Fig.~4(a)), $\tau_{P \rightarrow AP}$ decreases exponentially, while
$\tau_{AP \rightarrow P}$ increases only slightly. Varying $H$ while
holding $I$ fixed (Fig.~4(b)), on the other hand,
decreases $\tau_{AP \rightarrow P}$ exponentially, while $\tau_{P
\rightarrow AP}$ increases much more slowly.
These differences provide independent evidence that the spin-transfer
effect and $H$ alter the switching
of the nanomagnet in largely independent ways. By tuning $I$ and
$H$ along the easy axis, both switching times can be shortened
until the nanomagnet is effectively
superparamagnetic.

	In summary, magnetic reversal
driven by spin-polarized currents exhibits statistical properties of
thermal activation over an energy barrier.  This might appear to
favor an effective-magnetic-field model \cite{heide,guittienne} over
the torque model \cite{berger,slon}.  However, our data show that the
spin-transfer effect acts in a fundamentally different way than in
the effective field models,
while features of the torque model provide
natural explanations for
(1)
the different dependence on $H$ and $I$ of
$\sigma_{I_c}^{\pm}$ and $\sigma_{H_c}$, (2) the shape of the $T\!=\!0$
stability diagram for
P and AP orientations, and (3) the
distinct difference between the effects of $I$ and $H$ on switching
times in the
telegraph-noise regime.  Our data do not rule out a small
effective-field contribution
in addition to the torque term \cite{clear,zhang}.

We thank Piet Brouwer, Xavier Waintal, and Mandar Deshmukh for 
discussions.  Funding
was provided by ARO (DAAD19-01-1-0541) and NSF through the Nanoscale 
Science and Engineering Initiative, the Cornell
Center for Materials Research, and the use of the National
Nanofabrication Users Network.

\end{multicols}


\begin{thebibliography}{99}
\bibitem{tcomment}  In \cite{wernsdorfer}, the parameters $U_0$,
$H_{c0}$, $\alpha_H$, and $\tau_0$ were determined by fitting the
dependence of $\langle H_c \rangle$ on $\dot{H}$ and $T$.  We
cannot complete the analogous analysis for our samples because during
thermal sweeps $\langle I_c \rangle$ sometimes undergoes jumps larger
than $\sigma_{I_c}$.  This limits the range of $T$ accessible to quantitative
studies. Statistical measurements at fixed $T$ are fully reproducible.
\end{thebibliography}
\end{document}